\documentclass[useAMS,usenatbib]{mn2e}
\usepackage{graphicx}
\usepackage{ifthen}
\usepackage{times}
\usepackage{amsmath}
\usepackage{amssymb,amsfonts}
\usepackage{verbatim}
\usepackage{psfig}
\usepackage{epsf}

\def\ltsima{$\; \buildrel < \over \sim \;$}
\def\simlt{\lower.5ex\hbox{\ltsima}}
\def\gtsima{$\; \buildrel > \over \sim \;$}
\def\simgt{\lower.5ex\hbox{\gtsima}}
\def\kms{{\rm\,km\,s^{-1}}}

\def\kpc{{\rm\,kpc}}

\def\UseFigs{1}

\def\s{\ifmmode \widetilde \else \~\fi}
\def\={\overline}

\def\spose#1{\hbox to 0pt{#1\hss}}

\def\lta{\mathrel{\spose{\lower 3pt\hbox{$\mathchar"218$}}
     \raise 2.0pt\hbox{$\mathchar"13C$}}}
\def\gta{\mathrel{\spose{\lower 3pt\hbox{$\mathchar"218$}}
     \raise 2.0pt\hbox{$\mathchar"13E$}}}
\def\Dt{\spose{\raise 1.5ex\hbox{\hskip3pt$\mathchar"201$}}}    % upper case
\def\dt{\spose{\raise 1.0ex\hbox{\hskip2pt$\mathchar"201$}}}    % lower case

\def\dotsfill{\leaders\hbox to 1em{\hss.\hss}\hfill}

\loadboldmathitalic 
\title[A radial velocity survey of low Galactic latitude structures: II.] 
{A radial velocity survey of low Galactic
latitude structures:\\
II. The Monoceros Ring behind the Canis Major dwarf galaxy}
\author[Blair Conn et al.] 
       {Blair C. Conn$^1$, Nicolas F. Martin$^{2,4}$, Geraint F. Lewis$^1$, Rodrigo A. Ibata$^2$,
\newauthor Michele Bellazzini$^3$ \& Mike J. Irwin$^4$\\
$^{1}$Institute of Astronomy, School of Physics, A29, University of Sydney, NSW 2006, Australia:\\
Email \tt{bconn@physics.usyd.edu.au}, \tt{gfl@physics.usyd.edu.au}\\
$^{2}$Observatoire de Strabourg, 11, rue de l'Universit\'e, F-67000,
       Strasbourg, France:\\
 Email \tt{ibata@astro.u-strasbg.fr}, \tt{martin@astro.u-strasbg.fr}\\
$^{3}$INAF - Osservatorio Astronomico di Bologna, Via Ranzani 1,
       40127, Bologna, Italy:\\
Email \tt{michele.bellazzini@bo.astro.it}\\
$^{4}$Institute of Astronomy, Madingley Road, Cambridge, CB3 0HA,
       U.K.:\\
 Email \tt{mike@ast.cam.ac.uk}}

\begin{document}
\date{\today \hspace{10pt}}

\pagerange{\pageref{firstpage}--\pageref{lastpage}} \pubyear{2005}

\def\LaTeX{L\kern-.36em\raise.3ex\hbox{a}\kern-.15em
    T\kern-.1667em\lower.7ex\hbox{E}\kern-.125emX}

\newtheorem{theorem}{Theorem}[section]

\label{firstpage}

\maketitle 
\begin{abstract} 
An AAT/2dF Spectrograph Survey of low Galactic latitudes targeting the
putative Canis Major dwarf galaxy, and the (possibly) associated tidal
debris of stars known as the Monoceros Ring, covering Galactic
coordinates 231.5$^{\circ}\!\!<\,${\it l} $<$ 247.5$^{\circ}$ and
-11.8$^{\circ}\!\!<${\it b}$<$-3.8$^{\circ}$, has revealed the presence of
the Monoceros Ring in the background of the Canis Major dwarf
galaxy. This detection resides at a galactocentric distance of
$\sim$18.9$\pm$0.3\kpc~ (13.5$\pm$0.3\kpc~ heliocentric),
exhibiting a velocity of $\sim$132.8$\pm$1.3 $\kms$ with a
dispersion of $\sim$22.7$\pm$1.7$\kms$; both of these comparable with
previous measurements of the Monoceros Ring in nearby fields. This
detection highlights the increasing complexity of structure being
revealed in recent surveys of the Milky Way thick disk and Halo.
\end{abstract}


\section{Introduction}
Galaxy formation based on a $\Lambda$CDM cosmology predicts larger
structures being formed from the accumulation of smaller systems
~\citep[e.g.][]{1978ApJ...225..357S,1978MNRAS.183..341W,1978MNRAS.184..185W,2003ApJ...597...21A,2003ApJ...591..499A}.
One of the core outcomes of this model is that the halos of
galaxies should be strewn with the debris from all these minor
mergers. The first discovery of such
a merger  within our own Milky Way, the Sagittarius dwarf
galaxy~\citep{1994Natur.370..194I}, demonstrated that such mergers do
occur and are in fact ongoing.  

Insight into the complex nature of the Galactic Halo has been in part
accomplished by recent all sky surveys, Sloan Digital Sky Survey (SDSS) and the Two Micron All Sky Survey
(2MASS). Investigating an overdensity of F-turnoff stars in the SDSS,
~\citet{2002ApJ...569..245N} discovered the presence of a stream of
stars suggestive of an equatorial accretion event around the Milky
Way. Unlike the Sagittarius dwarf galaxy which is on a polar orbit,
this accretion event may have implications on the formation of key
Galactic structures, such as the thick disk. The equatorial orbit of the stream
was investigated by ~\citet{2003ApJ...594L.119C} who targeted 2MASS
selected M-giant stars in the range {\it l} = 150$^\circ$ -
240$^\circ$, although primarily in the Northern hemisphere, finding
that a simple circular orbit model with a galactocentric radius of
18\kpc~ and velocity of $\sim$220$\kms$ best fit the data. The velocity
dispersion of their sample was 20$\pm$4$\kms$ which excludes it from
being of Galactic origins. As more detections
of this structure were made
~\citep[][]{2004ApJ...605..575Y,2003MNRAS.340L..21I,2003ApJ...594L.115R}
it became clear that, as proposed by ~\citet{2002ApJ...569..245N},
this was indeed another on-going accretion event. The search for the progenitor of the stream in the
2MASS catalog led ~\citet{2004MNRAS.348...12M} to uncover the Canis
Major dwarf galaxy ({\it l},{\it b}) = $\sim$(240$^\circ$,-8$^\circ$).

Because of the low density of stream material on the sky it has been
necessary to employ the use of Wide-Field Cameras to look deeply into
Milky Way's thick disk and Halo, not only confirming and enhancing
previous detections but probing new regions leading to more detections of the Monoceros Ring
and the Canis Major dwarf being made
~\citep{2004MNRAS.348...12M,2004MNRAS.354.1263B,Conn2005a,2005Martinez}.  Detections of
an additional structure in Triangulum-Andromedae
~\citep{2004ApJ...615..732R,2004ApJ...615..738M} are revealing the increasing
complexity of accretion events in the outer regions of the Galaxy. Current
debate on these detections centres on whether these represent separate
accretion events or are the product of a single ongoing merger.  The
location of the progenitor is still contentious but this paper
will refer to the Canis Major overdensity as the Canis Major dwarf
galaxy (CMa), following the conclusions of ~\citet{2005Martinez} and ~\citet{2005Martin}. 

While the positions of the streams on the sky are
highly accurate, the distance estimates to the stars are less so, 
requiring the need for radial velocity kinematic measurements to
increase and characterize the known properties of the streams. Distances are then estimated
through photometric parallaxes as
described in \citet{2004MNRAS.355L..33M}. To date, several kinematic surveys of
the Monoceros Stream have been
undertaken using the data of 2MASS and SDSS
~\citep{2004ApJ...605..575Y,2003ApJ...594L.119C,2004ApJ...615..732R}.
These have confirmed a velocity gradient with Galactic longitude and
also a low velocity dispersion of $\sigma_{(v_r)} \sim $20$\kms$
for the Monoceros Ring.

While most models of equatorial accretion onto disk-like galaxies
~\citep[e.g.][]{2005ApJ...626..128P, 2005Martin} predict the presence of multiply
wrapped streams, no detections provide conclusive evidence supporting
this hypothesis. This is most likely due to the pencil beam nature of most fields with regard the entire
structure on the sky and given that confirmed detections of the
stream are limited to galactic longitudes ($120^{\circ}\!\!<{\it
  l}<240^{\circ}$) and significant areas of this structure are yet to
be sampled.  This new detection presented here is consistent with
multiply wrapped streams.

\section{Observations and reduction}\label{obs}
In April 2004, a survey was undertaken with the 2dF (Two-degree
Field) multi-fibre spectrograph on the Anglo-Australian Telescope (AAT)
with two aims: firstly, to determine the kinematics of the Canis
Major dwarf, and secondly to try and locate the stars belonging to
the Monoceros Ring through kinematic constraints.  Out of $\sim$15
fields, eight were dedicated to this purpose, four focusing solely on
the Canis Major region.  This paper will focus
primarily on results pertaining to the detection of the Monoceros
Ring in the Canis Major dwarf region, the same area studied by
\citet{2005Martin}; hereafter Paper I.

The 2dF instrument is capable of measuring velocities of $\sim$400
stars simultaneously within a two degree field of view. The field is
divided between two spectrographs, the first using a 1200V grating
(4600-5600\AA~ at 1\AA/pixel); the second using a 1200R grating
(8000-9000\AA~ at 1\AA/pixel). The configuration on the second spectrograph
was chosen to target Red Clump stars in the distance range 5-8 \kpc,
these stars fall outside the distance range expected for the Monoceros
Ring and thus are not discussed in this paper. (Paper I discusses the
results of the survey in the region $<$10 \kpc.)

The Red Giant Branch (RGB) stars in this survey were chosen from Sample A of
\citet{2004MNRAS.348...12M}. Using the photometric parallax technique of
~\citet{2003ApJ...599.1082M} to determine the distances, the stars in
the correct distance range for the Monoceros Ring can then be easily
selected. Determining the exact error on the distance using this
technique is difficult and so for these observations a conservative
estimate of $\pm$1 \kpc~ for these stars has been assumed. The J,H,K
magnitudes from the 2MASS catalogue for these stars have been
de-reddened using the dust maps of ~\citet{1998ApJ...500..525S} and
the asymptotic correction of ~\citet{2000AJ....120.2065B}. 

\subsection{Reducing 2dF spectra}\label{reduction}
Data reduction of the 2dF spectra involves using the 2dFDR package
~\citep{1996ASPC..101..195T} which applies the flat field correction,
sky subtraction and extracts the individual spectra from each fibre off the CCD. Due
to an asymmetry in the Line Spread Function (LSF) of the spectra a
{\it custom-made} reduction pipeline was constructed to minimise the errors
this produced when comparing with radial velocity standard stars; a
detailed presentation of the pipeline can be found in
~\citet{2005bMartin}. This pipeline limits the systematic offsets in
the radial velocity due to the LSF to $\pm$5$\kms$, this value is not
included in the errors stated. Determining
the radial velocities, $v_r$, using several artificial standard star templates and taking
a weighted average of the solutions avoids any systematic offset that may result
from a peculiarity in one of the templates ~\citep{2005Martin,2005bMartin}. Stars
then with $\sigma_{v} >$ 5$\kms$ (See Equation 2 of Paper I) are removed from the sample to avoid
contaminating the results with poor determination of radial
velocity. Dust extinction in this region typically is E(B-V)$<$0.4
~\citep[taken from][]{1998ApJ...500..525S} and only the field closest to the plane
has higher extinction. There has been no significant impact on the
2MASS colours due to extinction.

\subsection{Determining the Fit and the Associated Error}\label{fits}
For each velocity and distance distribution, a Gaussian has been
fitted to the profile using an {\tt amoeba} routine ~\citep{1992nrfa.book.....P},
obtaining a characteristic velocity, velocity dispersion, distance and
distance dispersion which best represents the data. The errors on the
fitted parameters were determined by bootstrap re-sampling and then
refitting the data. The re-sampling was undertaken using the
errors of each quantity namely the conservative $\pm$1 \kpc~ for the
distance to each star and the root-mean-square value of the dispersion. The errors cited then, are
the standard deviations of the fit determined using this method. There
is no appreciable difference between the galactocentric and
heliocentric errors in this field. The velocity errors also have an
additional $\pm5 \kms$ systematic error on top of the stated errors.

\section{Results}
\subsection{The Monoceros Ring behind the Canis Major dwarf}\label{results}
\begin{figure*}
\ifthenelse{\UseFigs=1}{
\includegraphics[angle=270,width=10cm]{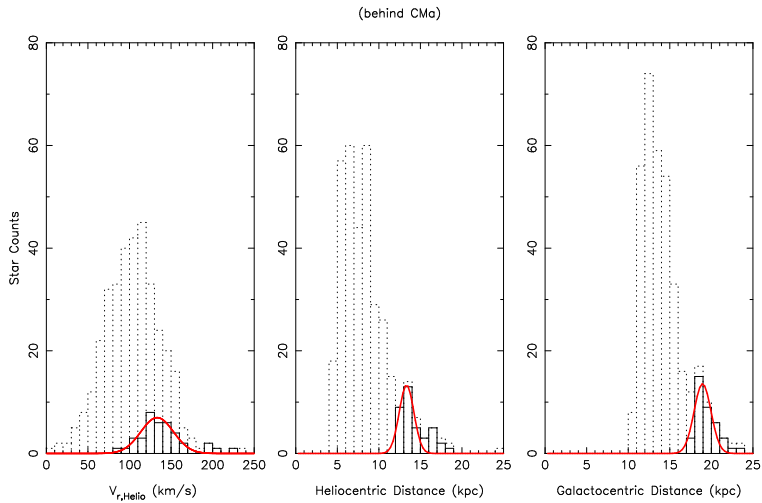} }{caption1}
\caption[]{Histogram of the 2dF RGB data in the range
  231.5$^{\circ}\!\!<$l$<$247.5$^{\circ}$ and
  -11.8$^{\circ}\!\!<$b$<$-3.8$^{\circ}$; with a heliocentric distance
  cut of 12\kpc$<D_{\odot}<$20\kpc~ and the following additional cuts of 0.9
  $<$(J-K)$_\circ\!\! <$ 1.3; ~0.561(J-K)$_\circ$ + 0.22 $<$
  (J-H)$_\circ$ $\!\!<$ 0.561(J-K)$_\circ$ + 0.36;~ K$_\circ$
  $\!\!\leq$ 13.0 and ~$v_r \!\!>$ 0$\kms$. This field constitutes the region
  behind the Canis Major overdensity revealing a detection of the
  Monoceros Ring at a galactocentric distance of 18.9$\pm$0.3 \kpc~
  and a dispersion of 0.8$\pm$0.4 \kpc~ (right panel) with a
  heliocentric distance of 13.5$\pm$0.3\kpc~ with similar dispersion
  (middle panel). The velocity profile fitted to this population is
  $v_{r}=$132.8$\pm$1.3$\kms$ with an internal dispersion of
  22.7$\pm$1.7$\kms$ (left panel), the error is accompanied by an
  additional systematic offset of $\pm5 \kms$.  The dashed lines shows
  the remaining data outside the distance and colour cut. For a full
  outline of the errors see $\S$\ref{fits} and Paper I.}{
\label{figCMa}}
\end{figure*}

The data in Figure~\ref{figCMa} was taken from six fields of the 2dF
spectrographic survey and shows a population of $\sim$38 RGB stars
behind the Canis Major overdensity. These stars are represented in the
solid histogram while the dashed histogram is the remaining stars in
that region. The stars have been selected using the following
criteria, $231.5^{\circ}\!\!<\!\!{\it l}\!\!<\!\!247.5^{\circ}$ and
-11.8$^{\circ}\!\!<$b$<$-3.8$^{\circ}$; 12\kpc$<\!\!D_{\odot}\!\!<$20\kpc~ and a
colour cut of 0.9 $<$(J-K)$_\circ$ $\!\!<$ 1.3; ~0.561(J-K)$_\circ$ + 0.22
$<$ (J-H)$_\circ$ $<$ 0.561(J-K)$_\circ$ + 0.36;~ K$_\circ$ $\!\!\leq$
13.0; with ~$v_r >$ 0$\kms$. This overdensity shows up very clearly in the
galactocentric distance histogram and given the distance range of
these stars makes them different from the foreground Canis Major stars
as presented in Paper I.  They form a coherent velocity and distance
structure similar to that of previous Monoceros Ring detections.
Figure~\ref{figCMa} shows the 2dF Data with a Gaussian fit of
$v_{r}=$132.8$\pm$1.3$\kms$ and internal dispersion of
22.7$\pm$1.7$\kms$, which is comparable to the dispersion as measured
by ~\citet{2004ApJ...605..575Y} from SDSS data. The galactocentric
distance estimates to this feature are 18.9$\pm$0.3 \kpc~ with a
dispersion of 0.8$\pm$0.4 \kpc. See $\S$\ref{fits} or Paper I for a
more complete breakdown on the errors, however note again that the
errors have an additional systematic offset of $\pm5 \kms$. To test
the contamination of local dwarfs which fall inside the selection
criteria for the sample stars due to poor extinction correction, the
data has been tested with a more stringent cutoff of
(J-K)$_\circ$$>$0.95. This still reproduces the detection presented
here suggesting the contamination level is low.

Although, as can be seen in Figure~\ref{figCMa}, there is no definitive
evidence for a {\it gap} between the Canis Major detection and the
Monoceros Ring detection this is most likely due to the large errors
in the distance rather than the structure being somewhat continuous out to
large galactocentric distances. The conservatively estimated error on
the distance is $\sim$1 \kpc, which will have the effect of
blurring any intrinsic gap between the two structures. Contamination
from local dwarfs although believed to be minimal will also impact on
the ability to resolve the stream out cleanly from the Canis Major
population. It should also be noted that depending on the formation scenario of the stream the
stars may or may not be neatly confined at one distance but rather
diffused over a larger volume.  Since the origins and evolution of both
structures is still being investigated, explaining the lack of a gap
in distance is at best speculative. With these stars being selected from the 2MASS catalogue a
measure of the completeness of the sample becomes possible. Applying
the colour cuts used on the 2dF data to the 2MASS catalogue reveals
399 stars in this region of which 364 have been observed making this
sample $\sim$91\% complete.

In Paper I, a northern comparison field at (240.0$^\circ$,
+8.8$^\circ$) is presented for the Canis Major
detection. Unfortunately, the number of RGB stars sampled out to large
Galactic radii is too low to form a useful comparison with the
detection presented in this paper.

\subsection{The Besan\c{c}on Synthetic Galaxy Model}\label{comparisons}
\begin{figure*}
\ifthenelse{\UseFigs=1}{
\includegraphics[angle=270,width=10cm]{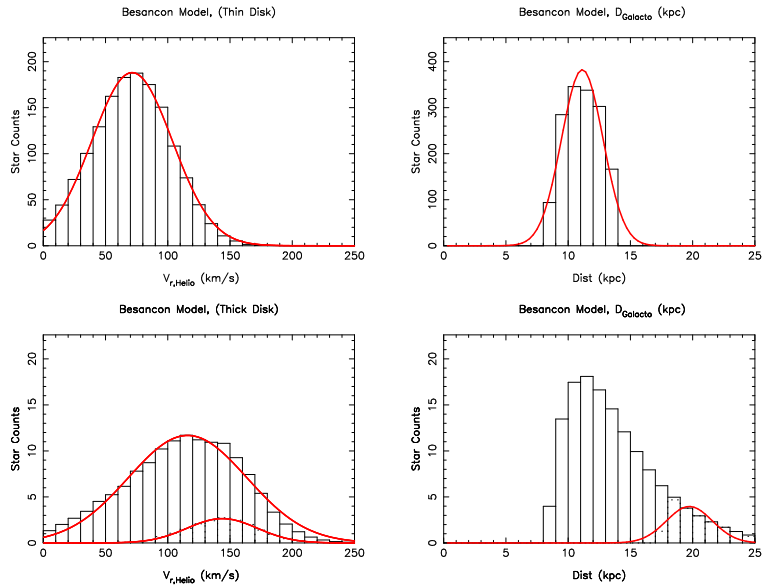} }{caption1}
\caption[]{Histogram of Besan\c{c}on data covering the same region of
  sky as the 2dF dataset; no distance cuts have been applied less than
  the 100$\kpc$ cutoff of the model. The distances are galactocentric
  to allow easy comparison with the 2dF data as shown in
  Figure~\ref{figCMa}. The top panels show the thin disk population
  (stars in the age range 5-7 Gyrs) in this field fitted with a
  Gaussian profile at a distance of $\sim$11.1$\kpc$ and a dispersion of
  $\sim$1.7$\kpc$. The velocity profile of the thin disk stars is $v_{r}$=
  $\sim$71.4$\kms$ with a dispersion of $\sim$32.3$\kms$. In the bottom panels are
  the thick disk stars (noting that these stars have a different
  density profile than the thin disk stars) they too can be fitted
  with a Gaussian profile and reside at a galactocentric distance of
  13.1$\kpc$~ with a dispersion of 5.0$\kpc$, the velocity and
  dispersion are found to be $v_{r}$= 115.9$\kms$ and 47.7
  $\kms$ respectively.  The smaller Gaussians represent those 15 stars in
  the same distance range as the MRi stars in Figure~\ref{figCMa} with
  $v_{r}$= 143.8$\kms$ and a dispersion of 28.2 $\kms$; galactocentric
  distance of 19.7$\kpc$~ with a dispersion of 1.9$\kpc$.}{
\label{figbes}}
\end{figure*}

Understanding what the Galactic contribution is in any given region of
sky is critical when attempting to interpret the results of this 2dF
survey. To investigate this, the synthetic galaxy model of
~\citet{2003A&A...409..523R} is used to find out what the model
predicts our Galactic contamination should be. This is done by selecting
the exact same regions of sky as observed in the 2dF survey.  Given
the tight constraints in colour and magnitude used to select the stars
in our survey (see Paper I) applying the same constraints to the
synthetic Galaxy model to provide an idea of the contamination we
should expect was undertaken.  In particular these are: 0.9 $<$
(J-K)$_\circ$ $\!\!<$ 1.3; ~0.561(J-K)$_\circ$ + 0.22 $<$ (J-H)$_\circ\!\!$
$<$ 0.561(J-K)$_\circ$ + 0.36;~ K$_\circ$ $\!\!\leq$ 13.0; ~$v_r\!\!>$
0$\kms.$ The cuts were applied to the model which extended to a
heliocentric distance of 100 \kpc~. 

To remove the presence of noise in the model, each field
was investigated using the online model selecting the {\it small field}
setting with a solid angle of 1000 deg$^2$. This has the effect of
providing a smoother distribution of the stars with the properties
determined by the centre of the field, rather than the {\it large
  field} option which takes into account changes in ({\it l,b})$^\circ$.  Each field is then
rescaled back to the size of the 2dF fields and rescaled to match the
completeness of the 2dF sample with regard the 2MASS catalogue. These
stars are presented in Figure~\ref{figbes} and in general have a much broader distribution
than could be attributed to any tidal debris with velocities which
typically peak well away from the 2dF detections. As expected, in the
distance range of the MRi there is no thin disk population and with the
thick disk only having $\sim$15 stars in this location;  halo stars are expected to
only have a very small presence in this field and are not included in
these results. The comparisons with the Besan\c{c}on model provide
good support for the validity of this detection with the usual precautions.

The approximately 15 stars which fall in the region of the MRi in the model
have the following properties, a velocity profile of $v_{r}$=
143.8$\kms$ and dispersion of 28.2 $\kms$; galactocentric distance
profile of 19.7$\kpc$~ with a dispersion of 1.9$\kpc$. A
Kolmogorov-Smirnoff test of the MRi stars from the 2dF sample and the
Besan\c{c}on stars finds a 4\% chance of the two being drawn from the same
distribution. This is a relatively high probability of coincidence
however a few extra factors need to be taken into account. While the
model contains thick disk stars with an additional warp component, the data contains thick disk stars, CMa stars and
MRi stars with a warp component mixed in. What effect does this have
on our comparison? Since a completeness factor has been derived from
the 2MASS catalogue and subsequently used to scale the Besan\c{c}on
model and given the 2MASS stars contains a significant proportion of CMa
stars then the numbers of thick disk stars must be being
overestimated.  This has the immediate effect of lowering the number
of stars that can be expected in the distance range of the MRi. The
true number of Galactic thick disk stars would be much less than the
$\sim$15 stars presented in Figure~\ref{figbes}, especially since most
of the stars as presented in Paper I seem to belong to the CMa
structure and that confusion over whether they represent the warp
seems to have been sufficently discounted.

\section{Discussion \& Conclusion}\label{discussion}

Comparisons with the Besan\c{c}on synthetic galaxy model
(Figure~\ref{figbes}) reveal that the peak velocity and
dispersion observed in this field cannot be associated with any
known Galactic component. The thick disk component in the same region as
the MRi has a higher velocity and a broader velocity dispersion which
discounts it as the cause of the overdensity measured in this field
(see Figure~\ref{figCMa}). This is supported by the finding that the
numbers of thick disk stars in the model are being overestimated
(after scaling) due to the presence of the CMa structure boosting the
numbers of stars in the region. Although difficult to quantify it is
expected that the numbers of Galactic thick disk stars should be much
less than the $\sim$15 stars as currently predicted. Investigating any similarities with
previous detections of the Monoceros Ring finds encouraging agreement
in the velocity dispersion estimates of ~\citet{2004ApJ...605..575Y}
and ~\citet{2003ApJ...594L.119C}, which have values of 13 - 24 $\kms$
and 20$\pm$4 $\kms$, respectively compared to 22.7$\pm$1.7 $\kms$ from
this paper. 

\begin{figure}
\ifthenelse{\UseFigs=1}{
\includegraphics[angle=270,width=8cm]{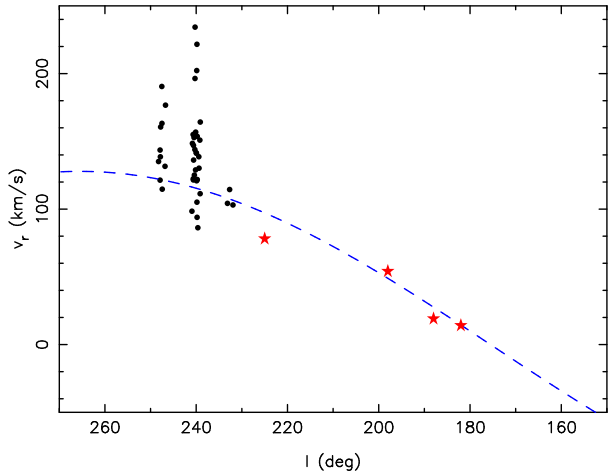} }{caption1}
\caption[]{Distribution of the target M giant stars compared to previous detections of the MRi. Stars from our sample are shown as filled circles whereas SDSS detections are plotted as stars. The dashed line corresponds to the best circular orbit model from ~\citet{2003ApJ...594L.119C} for a population orbiting the Milky Way at D$_{GC}$=18 \kpc~ with a rotational velocity of v$_{rot}$=220 $\kms$}{
\label{figcrane}}
\end{figure}

The best model of the MRi with which to compare our detection is the
~\citet{2003ApJ...594L.119C} circular orbit model since we target similar stars
(M giant stars extracted from 2MASS) and since they reach a Galactic
longitude of l$\sim$220 in the Southern hemisphere and l$\sim$240 degrees in the
Northern hemisphere. Moreover, as has already been noticed, the
velocity dispersion of our sample ($\sigma$ = 22.7$\pm$1.7 $\kms$) is statistically
similar to the ~\citet{2003ApJ...594L.119C} value (20$\pm$4
$\kms$). It can be seen on Figure~\ref{figcrane}, where our targets
are plotted as filled circles, that the velocity of our sample is only
slightly  higher than that expected by the circular model. Such a
small discrepancy is not unexpected since the MRi population certainly
does not follow a perfectly circular orbit (as is hinted by detections at different
galactocentric distances within the 14$<$D$_{GC}$$<$20 \kpc~ distance
range). Moreover, a closer look at Figure 2 of ~\citet{2003ApJ...594L.119C}
reveals the stars at high longitude in their sample (l$>\sim$220 deg) also
tend to have a radial velocity 20-30 $\kms$ higher than the circular
orbit model. Thus, our detection of the Monoceros Ring is consistent
with the current knowledge of this outer disk structure.

Equatorial accretion models predict the presence of wrapped tidal arms
and this detection provides a tentative confirmation of such a
scenario. A plausible alternative is that both the foreground Canis
Major detection and the background Monoceros Ring detection represent
separate accretion events.  However, such a situation complicates the
picture with the requirement of two progenitor systems hidden
somewhere in the Milky Way. Hence, we conclude that the probable
scenario is that the Canis Major dwarf and the Monoceros Ring
represent a single accretion event and this new detection is of a wrapped
tidal arm.

%\section{Conclusion}\label{conclusion}
\section{Acknowledgements}
BCC's thanks go to the Anglo-Australian Observatory and
the ANU Lodge for the technical support and hospitality received
during his stay.  GFL acknowledges the support of the Discovery
Project grant DP0343508 and would like to thank Little Britain and Trish, you
know Trish,.. Trish Trish, well anyway she's got nuffin to do wiv it, so Shyap!  
NFM acknowledges support from a Marie Curie Early Stage Research
Training Fellowship under contract MEST-CT-2004-504604.

\newcommand{\mnras}{MNRAS}
\newcommand{\nat}{Nature}
\newcommand{\araa}{ARAA}
\newcommand{\aj}{AJ}
\newcommand{\apj}{ApJ}
\newcommand{\apjl}{ApJ}
\newcommand{\apjs}{ApJSupp}
\newcommand{\aap}{A\&A}
\newcommand{\aaps}{A\&ASupp}
\newcommand{\pasp}{PASP}
\newcommand{\pasa}{PASA}

\label{lastpage}
\end{document}